\documentclass[reprint, amsmath, amssymb, aps, prl,]{revtex4-1}
\usepackage{amsmath}
\usepackage{graphicx}
\usepackage{dcolumn}
\usepackage{bm}
\usepackage{braket}
\usepackage{hyperref}
\hypersetup{colorlinks=true, citecolor=blue, urlcolor=blue, linkcolor=blue}

\begin{document}

\preprint{APS/123-QED}

\title{Deterministic Generation of Loss-Tolerant Photonic Cluster States with a Single Quantum Emitter}
\author{Yuan Zhan}
\author{Shuo Sun}
\email{shuosun@jila.colorado.edu}
\affiliation{JILA and Department of Physics, University of Colorado, Boulder, CO 80309, USA}

\date{\today}

\begin{abstract}
A photonic cluster state with a tree-type entanglement structure constitutes an efficient resource for quantum error correction of photon loss. But the generation of a tree cluster state with an arbitrary size is notoriously difficult. Here, we propose a protocol to deterministically generate photonic tree states of arbitrary size by using only a single quantum emitter. Photonic entanglement is established through both emission and re-scattering from the same emitter, enabling fast and resource-efficient entanglement generation. The same protocol can also be extended to generate more general tree-type entangled states.
\end{abstract}

\maketitle

Photons are unique carriers of quantum information. They have multiple degrees of freedom that can be employed to carry either qubits or qudits. In addition, photons are immune from thermal noise at room temperature, capable of long-distance transmission, and barely interact with each other. These properties make them ideal for quantum communication~\cite{RevModPhys.83.33,Northup:2014aa,RevModPhys.92.025002} and quantum networking~\cite{Kimble:2008aa,Wehner:2018aa}. While the lack of photon-photon interactions makes them less appealing in gate-based quantum computing, optical quantum computers can be constructed using a cluster state based model, known as ``measurement-based quantum computation''~\cite{PhysRevLett.86.5188,PhysRevLett.93.040503}. This model offers tremendous advantages for optical implementations since high-fidelity single-qubit gates and detectors can be realized with mature photonic devices~\cite{RevModPhys.79.135,OBrien:2007aa}.

A major obstacle in all these applications is the loss of photons during transmission, either in a quantum communication channel or in a delay line of an optical quantum computer. Imperfect quantum efficiency of single photon detectors can also be accounted as loss. Photon loss poses an exponential tradeoff between the rate and distance of repeaterless quantum communication in optical fibers~\cite{Pirandola:2017aa}, as well as a fundamental limit on the scalability of an optical quantum computer. It is thus essential to develop resource-efficient error correction methods that can deal with loss fault-tolerantly~\cite{PhysRevA.71.042323,PhysRevA.79.032325,PhysRevLett.112.250501}.

One such approach is to encode a qubit in a highly-entangled multi-photon cluster state, such as a tree cluster state~\cite{PhysRevLett.97.120501,PhysRevLett.100.060502}. The built-in redundancy in the tree-structure entanglement enables indirect measurement of a qubit even when a subset of the photons in the tree is lost~\cite{PhysRevLett.97.120501}. Unfortunately, generating a multi-photon entangled state with such a complicated entanglement structure is extremely challenging. Standard approaches rely on pairwise fusion gates to grow entangled photon pairs into a multi-photon cluster state~\cite{PhysRevLett.95.010501,Lu:2008aa,Azuma:2015aa,PhysRevA.95.012304}. The probabilistic nature of the fusion gates leads to a tremendous overhead on the required resources and a slow generation rate ($\sim$mHz for 12-photon entanglement with state-of-the-art experiments~\cite{PhysRevLett.121.250505}). To overcome this challenge, Lindner and Rudolph presented a deterministic protocol to generate multi-photon entangled states through sequential emission of photons from a single quantum emitter~\cite{PhysRevLett.103.113602}, but photons emitted in this process can only be entangled in the form of a one-dimensional chain. Inspired by this work, Buterakos et al. proposed a protocol that can sequentially emit photons into a repeater graph state with the help of an ancillary matter qubit~\cite{PhysRevX.7.041023}. While this protocol can be extended to generate a tree cluster state as shown in the same work, it requires as many ancillary matter qubits as the depth of the tree, along with the capability to perform two-qubit entangling gates between the quantum emitter and all the ancillary qubits. This demanding requirement limits the scale of the tree state that one can generate experimentally. In addition, since the entangling operation between the ancillary matter qubit and the quantum emitter is typically much slower than optical processes, the large number of entangling gates significantly reduces the generation rate of the tree state and the repeater graph state. In fact, it has been recently shown that the slow entangling operation between the matter qubits is the dominant limiting factor for the performance of the cluster state based all-optical quantum repeaters~\cite{hilaire2020resource}.

In this Letter, we propose a new protocol to deterministically generate a photonic tree state using only a single quantum emitter. The emitter is strongly coupled to a chiral waveguide that has a mirror at one end implementing a delayed feedback. The entanglement structure is thus established through both sequential emission of photons from the quantum emitter and re-scattering of photons following the delayed feedback, enabling fast and resource-efficient generation of the tree cluster states. While a similar scheme has been proposed to generate projected entangled pair states~\cite{Pichler:2017aa}, our proposal for the first time shows the capability to generate complex aperiodic entanglement structures with a simple delayed feedback. We also analyze our protocol under realistic error models and show that the protocol is robust against typical errors associated with its potential experimental platforms. Our proposal paves the way towards the realization of all-optical quantum repeaters~\cite{PhysRevA.79.032325,PhysRevLett.112.250501,Azuma:2015aa} and loss-tolerant optical quantum processors~\cite{PhysRevA.71.042323}.

\begin{figure}[tb]
\centering
\includegraphics[width=1.0\columnwidth]{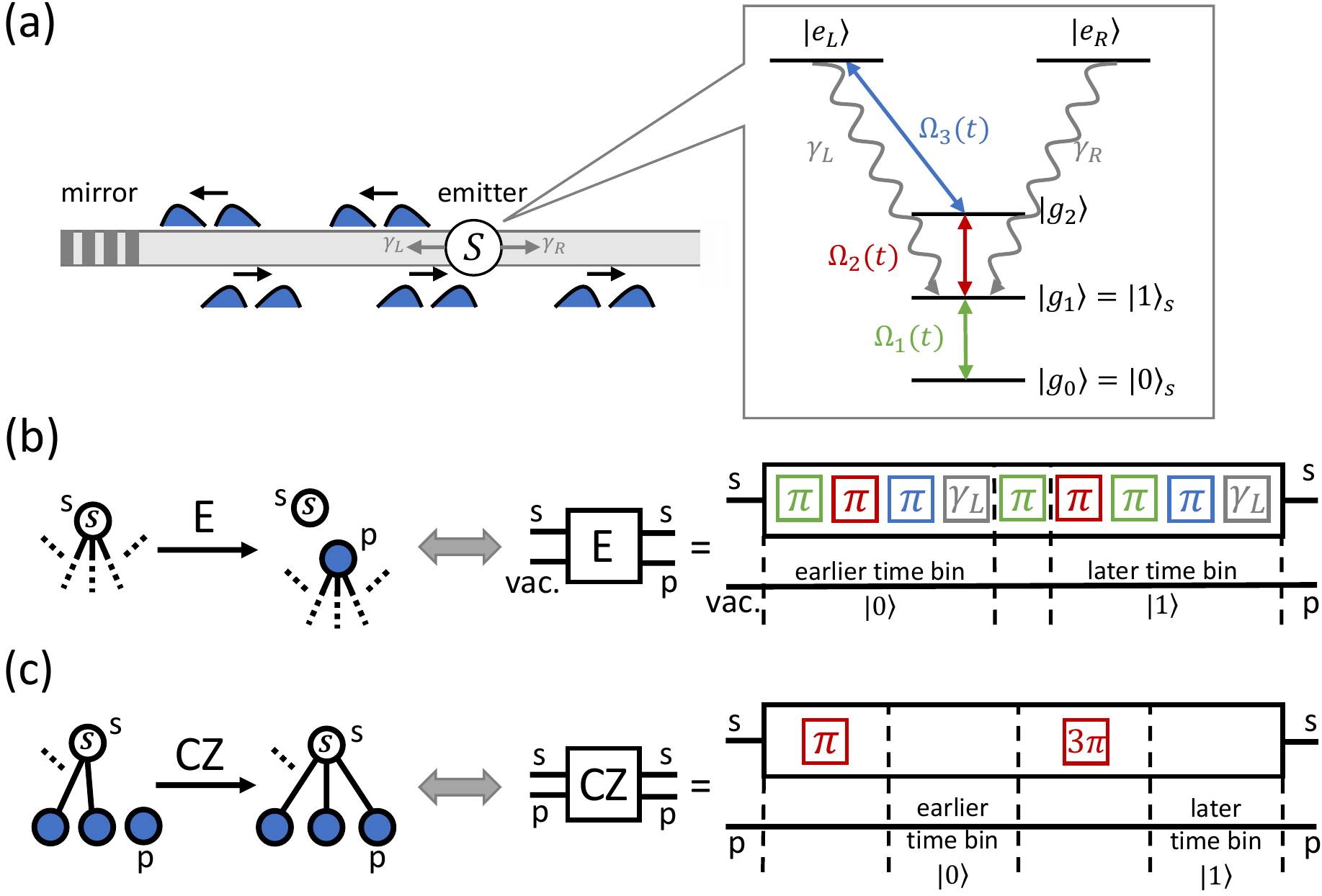}
\caption{(a) The schematic setup to generate a photonic tree cluster state with an arbitrary size. The inset shows the energy-level structure of the quantum emitter. (b, c) The left panels illustrate the effects of the E gate (b) and the CZ gate (c) applied on the joint emitter-photon quantum state. The right panels show the pulse sequences required to implement the E gate (b) and the CZ gate (c). The color of each block in the pulse sequence is used to indicate the optical transition [also color labeled in panel (a)] to which the rotation pulse is applied.}
\label{fig1}
\end{figure}

Figure~\ref{fig1}(a) shows the schematic setup we propose to generate the photonic tree state. It consists of a single quantum emitter (labeled $S$), a chiral waveguide, and a distant mirror placed at one end of the waveguide. For concreteness, we assume the emitter has an energy level structure as shown in the inset of Fig.~\ref{fig1}(a). It consists of three meta-stable ground states, labeled as $\ket{g_0}$, $\ket{g_1}$, and $\ket{g_2}$.  The states $\ket{g_0}$ and $\ket{g_1}$ form a stable qubit ($\ket{0}_s\equiv \ket{g_0}$ and $\ket{1}_s\equiv \ket{g_1}$), which can be coherently manipulated with a classical field $\Omega_1(t)$. The state $\ket{g_2}$ serves as an ancillary memory state used in the generation of time-bin encoded photons as will be explained next. The population of this state can be coherently prepared from the state $\ket{g_1}$ with another classical field $\Omega_2(t)$. The quantum emitter also consists of two optically excited states $\ket{e_L}$ and $\ket{e_R}$. Both excited states can decay into the ground state $\ket{g_1}$ while emitting a photon into the waveguide. We assume the couplings between the emitter and the waveguide are chiral, such that the transitions $\ket{g_1}\leftrightarrow\ket{e_L}$ and $\ket{g_1}\leftrightarrow\ket{e_R}$ couple only to the left- and right-propagating modes of the waveguide, respectively. Such chiral couplings have been experimentally demonstrated across a number of atomic systems~\cite{Lodahl:2017aa}. We assume that we can drive the emitter into the excited state $\ket{e_L}$ from the ancillary ground state $\ket{g_2}$ with an optical laser $\Omega_3(t)$.

We encode each photonic qubit in the time-bin basis consisting of two possible temporal modes well separated from each other. Specifically, we denote the presence of a photon in the earlier and later temporal modes as $\ket{0}_p$ and $\ket{1}_p$ respectively. The time-bin encoding is uniquely suitable for long-distance quantum communication~\cite{PhysRevLett.82.2594}, as it is robust to depolarization errors and also allows for the detection of photon loss.

We first introduce the two elementary gates required to implement our protocol, the E gate and the controlled-Z (CZ) gate. The left panel in Fig.~\ref{fig1}(b) illustrates the action of the E gate on the joint emitter-photon quantum state. An E gate generates a new photon that inherits the state of the quantum emitter, while resetting the state of the quantum emitter to $\ket{1}_s$. Mathematically, the transformation of the E gate can be written as $(\alpha\ket{0}_s\ket{\psi_0}_r+\beta\ket{1}_s\ket{\psi_1}_r)\ket{\text{vacuum}}_{p}\rightarrow(\alpha\ket{0}_{p}\ket{\psi_0}_r+\beta\ket{1}_{p}\ket{\psi_1}_r)\ket{1}_s$, where $s$, $r$, and $p$ represent the states of the emitter, the rest of the photons that are already emitted which may be entangled with the emitter, and the newly-generated photon respectively. The right panel in Fig.~\ref{fig1}(b) shows the pulse sequence required to implement the E gate. By successively applying three $\pi$-pulses of $\Omega_1(t)$, $\Omega_2(t)$ and $\Omega_3(t)$, the emitter can be excited to state $\ket{e_L}$ and emits a left-propagating photon into the earlier time-bin if it is initially in state $\ket{g_0}$, while populated to $\ket{g_0}$ if initially in $\ket{g_1}$. We next apply a $\pi$-pulse of $\Omega_1(t)$ to swap $\ket{g_0}$ and $\ket{g_1}$, and repeat the process to generate a left-propagating photon into the later time-bin if the emitter is initially in state $\ket{g_1}$. Another $\pi$-pulse of $\Omega_1(t)$ is used to make sure that the emitter is reset to $\ket{g_1}$ regardless of its initial state. Figure~\ref{fig1}(c) shows the action of the CZ gate on the joint emitter-photon quantum state, along with the required pulse sequence for its realization. The CZ gate is applied between the emitter and a photon reflected from the mirror. To implement the CZ gate, we apply a $\pi$-pulse of $\Omega_2(t)$ before the earlier time-bin, and another $3\pi$ pulse of $\Omega_2(t)$ in the middle of the earlier and later time-bins. Therefore, if the photon is in the state $\ket{1}_p$ (later time-bin), it will pick up a $\pi$ phase shift if the emitter is initially in the state $\ket{g_1}$ due to the strong coupling between the transition $\ket{g_1}\leftrightarrow\ket{e_R}$ and the right-propagating mode of the waveguide~\cite{PhysRevLett.96.153601,Fan:2010aa}, but no phase shift if the emitter is initially in the state $\ket{g_0}$. A photon in state $\ket{0}_p$ (earlier time-bin) will always transmit with no phase shift since the emitter can only be in state $\ket{g_0}$ or $\ket{g_2}$ during the earlier time-bin. The $3\pi$-rotation in the second $\Omega_2(t)$ pulse, instead of a $\pi$-rotation, avoids accumulation of a $\pi$ geometric phase between the emitter states $\ket{g_0}$ and $\ket{g_1}$.

\begin{figure}[tb]
\centering
\includegraphics[width=1.0\columnwidth]{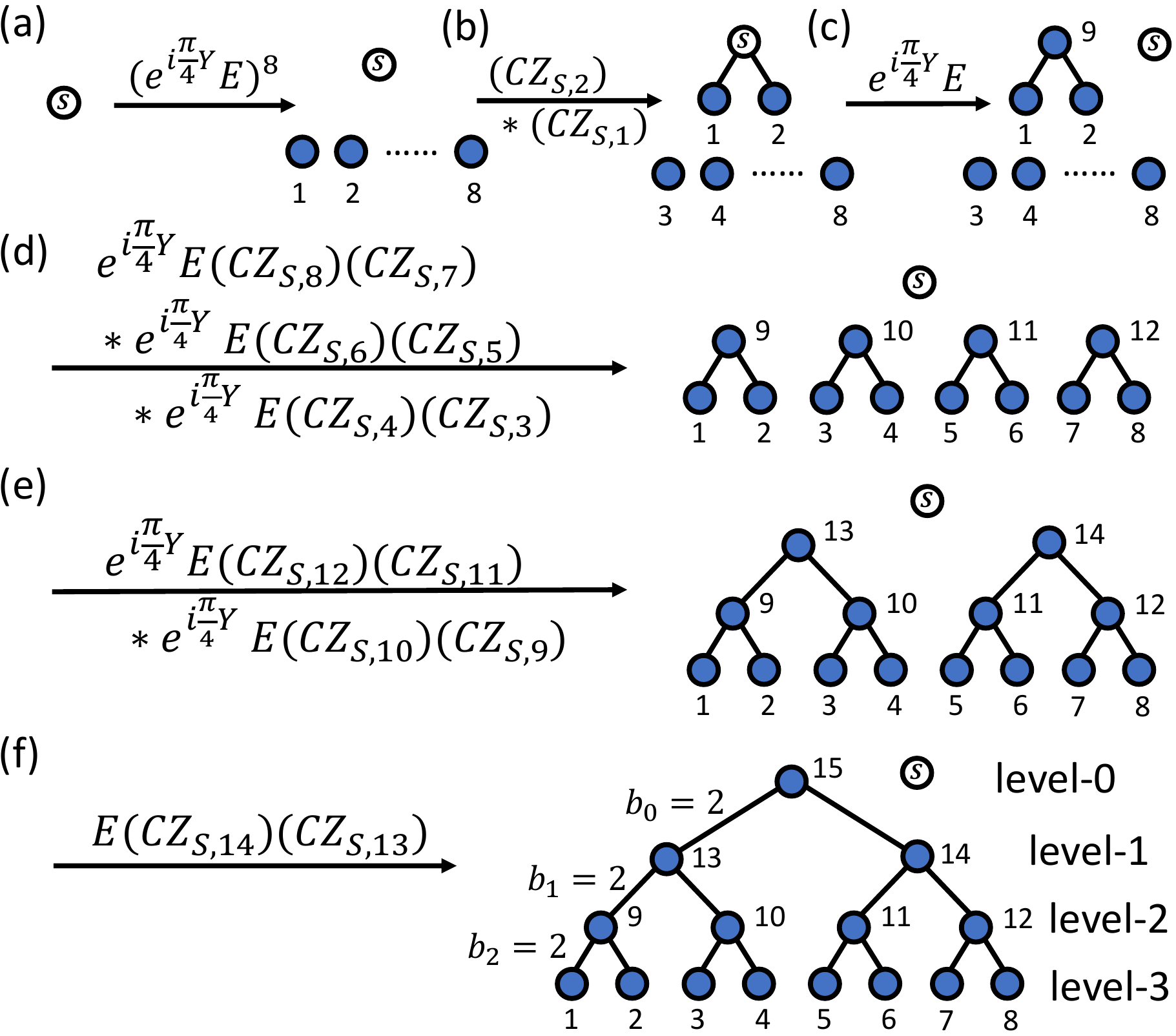}
\caption{Graph representation of the procedure for generating a tree with branching parameters $b_0=b_1=b_2=2$.}
\label{fig2}
\end{figure}

To provide an intuitive understanding, we first describe our protocol using an example of a tree shown in Fig.~\ref{fig2}(f), which has a depth of 3 and branching parameters of $b_0=b_1=b_2=2$. However, it should be noted that our protocol applies to any arbitrary tree structures and sizes. Here we define the branching parameter $b_i$ as the number of leaf nodes connected with a node at level-$i$, where the level number $i$ is defined as the number of edges on the path from the node of interest to the root node of the tree. Our protocol generates photons from the bottom of the tree to the top, as shown by the graph representation of the procedure in Fig.~\ref{fig2}. We start with the emitter prepared in state $\ket{+}_s=\frac{1}{\sqrt{2}}(\ket{0}_s+\ket{1}_s)$. By continuously applying the E gate and a ($-\pi/2$) spin rotation along the $y$-axis of the Bloch sphere for 8 times, we generate 8 photons that are all in the state $\ket{+}_p=\frac{1}{\sqrt{2}}(\ket{0}_p+\ket{1}_p)$ [see Fig.~\ref{fig2}(a)]. These photons will constitute the bottom layer of the tree. The 8 photons will travel sequentially in the left-propagating mode of the waveguide until they are reflected by the mirror. For each reflected photon, we apply a CZ gate when the photon arrives at the emitter. Since both the emitter and the photon are in the superposition state, the CZ gate entangles the emitter and the photon. Therefore, after the first two photons pass through the emitter, the emitter will be entangled with both photons as shown in Fig.~\ref{fig2}(b). Before the 3rd photon arrives at the emitter, we apply an E gate to generate a new photon (the 9th photon) into the left-propagating mode of the waveguide. This E gate will transfer the state of the emitter into the 9th photon and reset the emitter to state $\ket{1}_s$. Thus the 9th photon becomes the parent node of the photons 1 and 2, and the emitter is detached from this sub-tree [see Fig.~\ref{fig2}(c)]. A follow-up ($-\pi/2$)-rotation along $y$-axis on the emitter will prepare the emitter back to the $\ket{+}_s$ state again. Repeating the same procedure for another three times will generate three more subtrees as shown in Fig.~\ref{fig2}(d). Up to now, the photons 1-8 have passed through the emitter in the right-propagating mode and will no longer interact with the emitter, whereas the photons 9-12 are in the left-propagation mode of the waveguide and will be reflected back to the emitter. Following the same procedure, we will again entangle the emitter with both the photons 9 and 10 through two sequential CZ gates applied when they arrive at the emitter, and transfer the emitter state into another newly generated photon (the 13th photon) through an E gate. Repeating this procedure one more time will generate two larger subtrees as shown in Fig.~\ref{fig2}(e). Lastly, we will repeat the same procedure for the last time and generate the root node of the tree using an E gate (photon 15), which also decouples the emitter from the whole tree.

We now formalize our protocol for generating a general tree state with a depth of $d$ and branching parameters $\{b_0,b_1,\cdots,b_{d-1}\}$. The sequence of operations can be described as follows:
\begin{widetext}
\begin{equation}
\begin{aligned}
\ket{\psi_{\text{tree}}}=\prod_{j=1}^d\left[\prod_{k=1}^{n_{d-j}}\left(e^{i\frac{\pi}{4}Y}E\prod_{l=1}^{b_{d-j}}CZ_{S,\sum_{m=0}^{j-1}n_{d+1-m}+(k-1)b_{d-j}+l}\right)\right]\left(e^{i\frac{\pi}{4}Y}E\right)^{n_d}\ket{+}_s,
\end{aligned}
\end{equation}
\end{widetext}
where $n_l=\prod_{i=0}^{l-1}b_i$ is the number of photons in the $l$-th level of the tree. $CZ_{S,i}$ represents a CZ gate applied on the emitter $S$ and the $i$-th generated photon.

While we described our protocol using a specific system consisting of a multi-level atom coupled to a chiral waveguide, it is worth noting that neither the specific atomic level structure nor the chiral coupling is essential to the realization of our protocol. For example, our protocol can be realized with a cavity QED device with a simple II-type and $\Lambda$-type atom, as described in the Supplemental Materials~\cite{Note1}. Such a cavity QED system has been realized by a number of atomic systems including trapped Rb atoms~\cite{RevModPhys.87.1379}, semiconductor quantum dots~\cite{RevModPhys.87.347,Sun:2016aa}, and diamond color centers~\cite{Sipahigil847}.

\begin{figure}[tb]
\centering
\includegraphics[width=1.0\columnwidth]{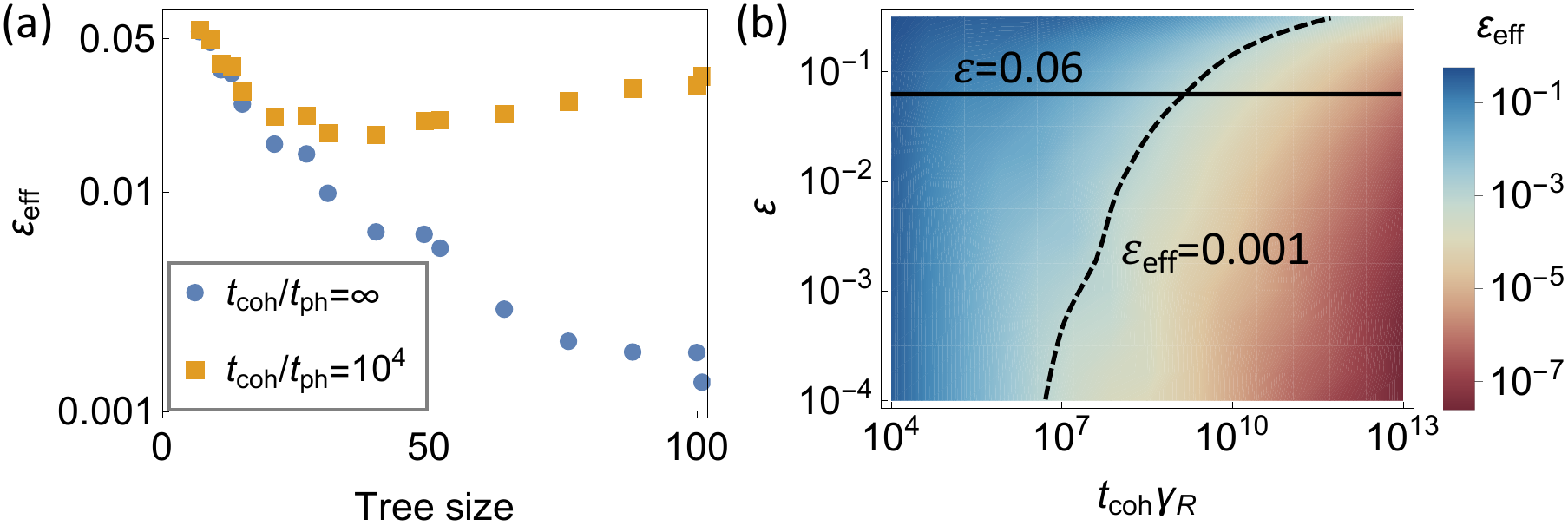}
\caption{(a) The effective error probability of the logic qubit encoded by the tree as a function of the tree size (the total number of photons in the tree). We assume a single-photon loss probability of $\varepsilon=0.1$ in this calculation. The blue circles represent the case where $t_{\text{coh}}\to\infty$, and the orange squares represent the case where $t_{\text{coh}}/t_{\text{ph}}=10^4$. (b) The optimized value of $\varepsilon_{\text{eff}}$ as a function of single photon loss probability and coherence-time-bandwidth product.}
\label{fig3}
\end{figure}

We now discuss the robustness of our protocol against typical errors during the tree state generation. One of the dominant errors in any photon generation process is the internal photon loss, which in our scheme may result from emission or rescattering of photons into the bath other than the waveguide mode due to the finite cooperativity, absorption during photon transmission in the waveguide, and partial reflection from the end mirror. While the external loss is quantum error correctable due to the tree-type encoding, the internal loss may lead to uncorrectable errors since it happens before the tree entanglement is fully established. Following the proof shown in Ref.~\cite{PhysRevLett.97.120501}, we can show that the internal loss can be indeed corrected in the same way as the external photon loss. This is fundamentally because the attempted CZ gate operates as an identity operation when a photon is lost internally. Therefore, our protocol is loss resilient as long as the total loss is below the quantum error correction threshold of 50\%~\cite{PhysRevLett.97.120501}. As an example, we consider a specific application of performing an arbitrary single-qubit measurement on the tree-encoded logic qubit. We define the effective error probability $\varepsilon_{\text{eff}}$ as the probability that this measurement yields an incorrect result. The blue circles in Fig.~\ref{fig3}(a) show the value of $\varepsilon_{\text{eff}}$ as a function of the total number of photons in the tree~\cite{Note1}, with a single-photon loss probability of $\varepsilon=0.1$. The effective error probability decreases exponentially with the tree size and can eventually approach 0 given a large enough tree.

While we can in principle overcome the photon loss by generating a large enough tree, in practice the size of the tree we can generate is limited by the finite coherence time of the emitter qubit. The orange squares in Fig.~\ref{fig3}(a) show the effective error probability $\varepsilon_{\text{eff}}$ when we consider a finite emitter coherence time $t_{\text{coh}}$. Here we set $t_{\text{coh}}/t_{\text{ph}}=10^4$~\cite{Note1}, where $t_{\text{ph}}$ is the time allocated to a single-photon qubit. We have kept the same single-photon loss rate of $\varepsilon=0.1$ in the calculation. As we can see from Fig.~\ref{fig3}(a), when the tree size is small, the effective error probability is nearly identical with the value under infinite emitter qubit coherence time (blue circles), since the probability of emitter decoherence during the tree-state generation is negligible. However, as we keep increasing the tree size, the effective error probability tapers off and starts to increase, indicating that error caused by the emitter decoherence during the tree-state generation starts to dominate. Thus, given a qubit coherence time and single-photon loss rate, there is an optimized value of $\varepsilon_{\text{eff}}$ that we can achieve by varying the tree size.

Figure~\ref{fig3}(b) shows the optimized value of $\varepsilon_{\text{eff}}$ at different system parameters~\cite{Note1}. In this calculation, we vary both the single-photon loss rate $\varepsilon$ and the emitter qubit coherence time $t_{\text{coh}}$, while assuming all other possible imperfections absent (see Supplemental Materials~\cite{Note1} for a fidelity and error analysis when accounting for more experimental imperfections). We normalize $t_{\text{coh}}$ in terms of the inverse of the bandwidth of the CZ gate, $\gamma_R$, which is the waveguide modified linewidth of transition $\ket{g_2}\leftrightarrow\ket{e_R}$. As we can see, a small single-photon loss rate and a large coherence-time-bandwidth product $t_{\text{coh}}\gamma_R$ are needed to achieve a small $\varepsilon_{\text{eff}}$. To qualitatively identify the useful regime of $\varepsilon_{\text{eff}}$, we consider a specific application of using tree states to implement one-way quantum repeaters~\cite{Azuma:2015aa,PhysRevA.95.012304,PhysRevX.10.021071}. We assume a realistic internal photon loss rate of 0.01, and we distribute the quantum repeater nodes one in every 1 km. This distance corresponds to an external photon loss rate of 0.05 in a telecom fiber, which is small enough compared with the threshold of 0.5 for loss correction, but large enough to ensure that dominant loss to be corrected is from the optical fiber. Since we have $10^3$ repeater nodes, to achieve a reasonable communication rate over 1000 km requires the effective error probability of each link to be less than $\sim 10^{-3}$. The dashed line in Fig.~\ref{fig3}(b) denotes the parameter regime where $\varepsilon_{\text{eff}}=10^{-3}$, and the solid line shows the condition where $\varepsilon=0.06$, corresponding to the total loss rate of a single photon. Thus, to achieve $\varepsilon_{\text{eff}}<10^{-3}$ while $\varepsilon>0.06$ requires a coherence-time-bandwidth product exceeding $10^9$. This parameter regime can be possibly achieved upon reasonable improvements by using a single silicon-vacancy color center coupled with a photonic crystal cavity ($\gamma_R\sim 2\pi\times 10$ GHz~\cite{Bhaskar:2020aa} and $t_{\text{coh}}\sim 10$ ms~\cite{PhysRevLett.119.223602}), or a single trapped atom strongly coupled with a fiber Fabry-Perot cavity ($\gamma_R\sim 2\pi\times 100$ MHz~\cite{Brekenfeld:2020aa} and $t_{\text{coh}}\sim 1$ s~\cite{RevModPhys.87.1379}). It may also be possible to use a strongly coupled quantum dot and nano-cavity ($\gamma_R\sim 2\pi\times 80$ GHz~\cite{Ota:2018aa}) to reach this parameter regime, if one can improve its spin coherence time to $\sim 2$ ms (currently it is $\sim 4$ $\mu$s~\cite{PhysRevB.97.241413}).

In conclusion, we have proposed and analyzed a protocol for deterministic generation of tree-type photonic cluster states using only a single quantum emitter. Our protocol can generate a tree state with an arbitrary size and depth without any probabilistic fusion gates or ancillary matter qubits, which significantly reduces the resource overhead required for the entanglement generation. The protocol is also robust to typical errors in realistic experiments and is within reach upon reasonable improvements of quantum photonics technologies. In addition, our scheme can be implemented with a variety of cavity QED systems using both free-space optics and integrated photonics~\cite{Note1}.

One of the most important features of our protocol is that it can be widely applicable to a large range of tree-type photonic cluster states that are useful for all-optical quantum repeaters, such as the repeater graph states~\cite{Azuma:2015aa,PhysRevA.95.012304} and the tree-encoded repeater graph states~\cite{PhysRevA.95.012304,PhysRevX.7.041023} (see Supplemental Materials for details~\cite{Note1}). Thus, an important future work is to perform quantitative rate-distance tradeoff and resource cost analysis of our scheme in the application of different all-optical quantum repeater protocols by accounting for all possible losses, bit-flips, and dephasing errors, and systematically compare its performance with existing cluster state generation schemes~\cite{PhysRevA.95.012304,hilaire2020resource,PhysRevX.10.021071}. Overall, our results constitute an important scheme for aperiodic 2D cluster state generation with feasible resources, and pave the way towards the realization of loss-tolerant one-way optical quantum computers~\cite{PhysRevA.71.042323} and all-optical quantum repeaters~\cite{PhysRevA.79.032325,PhysRevLett.112.250501,Azuma:2015aa}.

We thank Paul Hilaire, Sophia Economou, and Edwin Barnes for fruitful discussions.

\nocite{*}

\providecommand{\noopsort}[1]{}\providecommand{\singleletter}[1]{#1}%

\end{document}


\title{Deterministic Generation of Loss-Tolerant Photonic Cluster States with a Single Quantum Emitter: Supplemental Materials}
\author{Yuan Zhan}
\author{Shuo Sun}
\affiliation{JILA and Department of Physics, University of Colorado, Boulder, CO 80309, USA}

\maketitle

\section{Alternative experimental implementations of the proposed protocol}

\begin{figure}[htb]
\centering
\includegraphics[width=0.7\columnwidth]{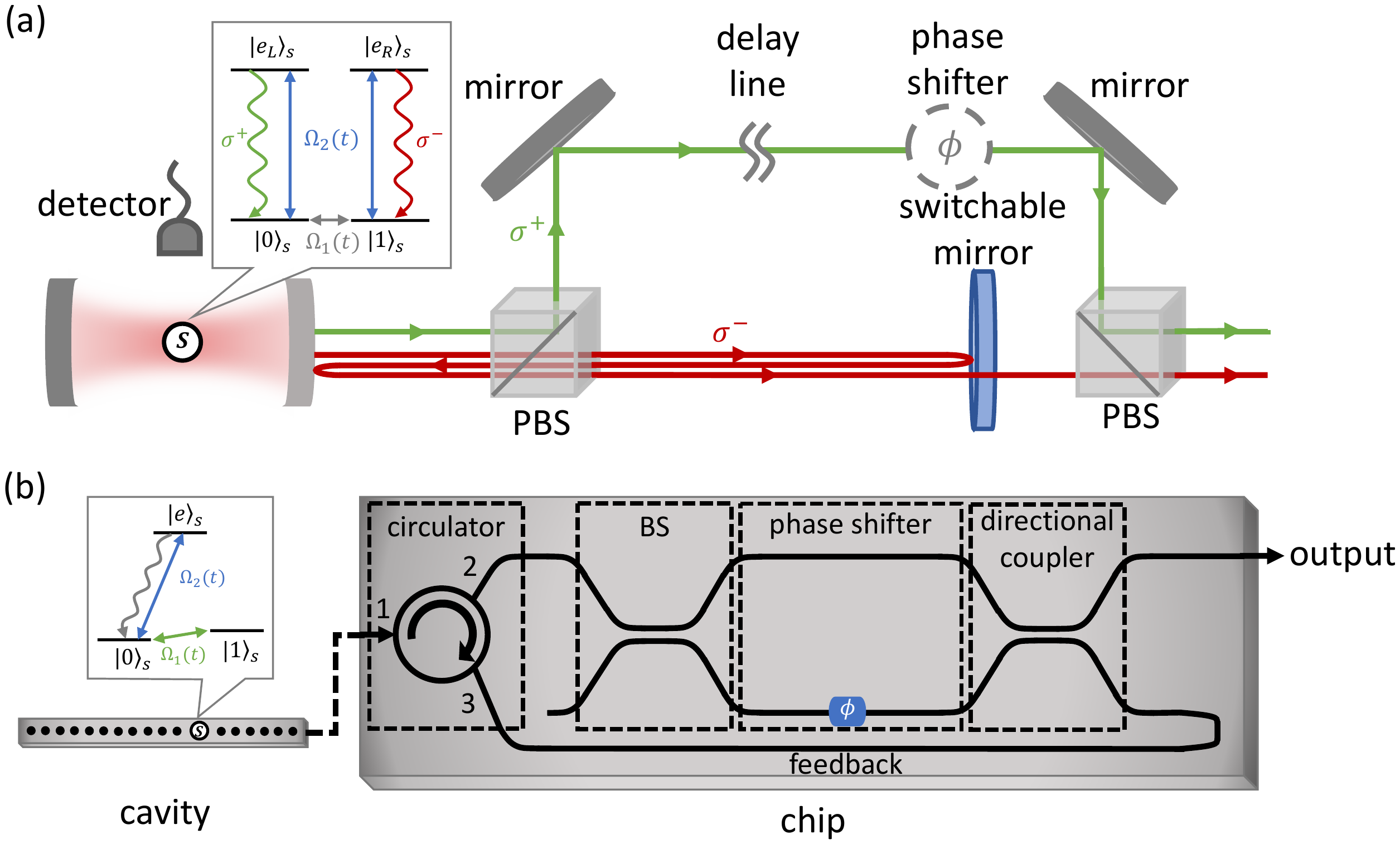}
\caption{Alternative experimental implementations of our protocol using a cavity QED device with an atom possessing a simple II- (a) or $\Lambda$-type (b) energy level structure. The delayed feedback is induced through either a switchable mirror in free-space optics (a) or phase-tunable Mach-Zehnder interferometers in integrated photonics (b).}
\label{cavity}
\end{figure}

In the main text, we describe our protocol using a specific system consisting of a multi-level atom coupled to a chiral waveguide. It is worth noting that neither the specific atomic level structure nor the chiral coupling is essential to the realization of our protocol. Our protocol can be realized with any coherent atom-photon interface possessing a large enough cooperativity, along with a delayed feedback. Figure~\ref{cavity} shows the schematic setup of two possible implementations of our protocol using a cavity QED device with an atom possessing a simple II- [Fig.~\ref{cavity}(a)] or $\Lambda$-type [Fig.~\ref{cavity}(b)] energy level structure. A reconfigurable optical element is used to implement the delayed feedback, which can be either a switchable mirror in free-space optics [Fig.~\ref{cavity}(a)] or phase-tunable Mach-Zehnder interferometers in integrated photonics [Fig.~\ref{cavity}(b)].

In the setup shown in Fig.~\ref{cavity}(a), an atom with a II-type energy level structure is coupled to a single-sided cavity. The atom has two ground states $\ket{0}_s$ and $\ket{1}_s$, and two excited states $\ket{e_L}_s$ and $\ket{e_R}_s$. Only transitions $\ket{0}_s\leftrightarrow\ket{e_L}_s$ and $\ket{1}_s\leftrightarrow\ket{e_R}_s$ are optically allowed. For concreteness of discussion, we assume that the polarizations of the two transitions are orthogonal, given by $\sigma^+$ and $\sigma^-$ respectively. In this case we can use the polarization basis to encode the photonic qubit ($\ket{0}_p\equiv \ket{\sigma^+}$ and $\ket{1}_p\equiv \ket{\sigma^-}$). Alternatively, we can also slightly detune the transitions $\ket{0}_s\leftrightarrow\ket{e_L}_s$ and $\ket{1}_s\leftrightarrow\ket{e_R}_s$, and encode the photonic qubit under the frequency-bin basis. We assume each of the two transitions couples to a mode of an optical cavity. In the case of the orthogonal polarization selection rule described above, the cavity needs to be polarization degenerate. To implement the E gate, we first pump the emitter using a $\pi$-pulse of $\Omega_2$(t) with a polarization $(\ket{\sigma^+}+\ket{\sigma^-})$, which drives the two transitions $\ket{0}_s\leftrightarrow\ket{e_L}_s$ and $\ket{1}_s\leftrightarrow\ket{e_R}_s$ simultaneously with an equal Rabi frequency. This operation transforms the emitter-photon system from the state $(\alpha\ket{0}_s\ket{\psi_0}_r+\beta\ket{1}_s\ket{\psi_1}_r)\ket{\text{vacuum}}_p$ to the state $(\alpha\ket{e_L}_s\ket{\psi_0}_r+\beta\ket{e_R}_s\ket{\psi_1}_r)\ket{\text{vacuum}}_p$, where $s$, $r$, and $p$ represent the states of the emitter, the rest of the photons that are already emitted, and the newly-generated photon respectively. Following the spontaneous emission, the state is transformed to $(\alpha\ket{0}_s\ket{0}_p\ket{\psi_0}_r+\beta\ket{1}_s\ket{1}_p\ket{\psi_1}_r)$. We next apply a $\sigma_x$ measurement on the emitter, which leaves the system in state $(\alpha\ket{0}_p\ket{\psi_0}_r+\beta\ket{1}_p\ket{\psi_1}_r)\ket{+}_s$ if the measurement outcome is $+$, while in state $(\alpha\ket{0}_p\ket{\psi_0}_r-\beta\ket{1}_p\ket{\psi_1}_r)\ket{-}_s$ if the measurement outcome is $-$. In the latter case, we can re-interpret the basis of the newly generated photon as $\ket{0}_p\equiv \ket{\sigma^+}$ and $\ket{1}_p\equiv -\ket{\sigma^-}$, so that the resulting state of the photon is always $(\alpha\ket{0}_p\ket{\psi_0}_r+\beta\ket{1}_p\ket{\psi_1}_r)$. We also need to apply an additional $\pi$ rotation on the emitter qubit to reset it back to the $\ket{+}_s$ state. These operations together realize the E gate, which generates a new photon that inherits the state of the emitter qubit. To implement the delayed feedback and the CZ gate, we use a polarizing beam splitter (PBS) to split the two polarization components of the photonic qubit to different spatial modes, and redirect only the $\ket{1}_p$ wavepacket back to the cavity for re-scattering by using a switchable mirror. Upon rescattering, the $\ket{1}_p$ photonic wavepacket picks up no phase shift if the emitter is in state $\ket{1}_s$, and a $\pi$ phase shift if the emitter is in state $\ket{0}_s$, which realizes the CZ gate between the photon and the emitter qubit. We switch off the mirror when the $\ket{1}_p$ component of the photonic qubit arrives for the second time, and recombine it with the $\ket{0}_p$ component at a second PBS before the photon exits the setup.

In the setup shown in Fig.~\ref{cavity}(b), an atom with a $\Lambda$-type energy level structure is coupled to a single-sided nanophotonic cavity. The atom has two ground states $\ket{0}_s$ and $\ket{1}_s$ and an excited state $\ket{e}_s$. We assume that only the transition $\ket{0}_s\leftrightarrow\ket{e}_s$ is optically allowed and couples to the cavity. We encode the photonic qubit under time-bin basis, where the earlier and later time bins are encoded as $\ket{0}_p$ and $\ket{1}_p$ respectively. To implement the E gate, we first pump the emitter with a $\pi$-pulse of $\Omega_2$(t), which transforms the emitter-photon state from $(\alpha\ket{0}_s\ket{\psi_0}_r+\beta\ket{1}_s\ket{\psi_1}_r)\ket{\text{vacuum}}_p$ to $(\alpha\ket{e}_s\ket{\psi_0}_r+\beta\ket{1}_s\ket{\psi_1}_r)\ket{\text{vacuum}}_p$. Following the spontaneous emission, the state is transformed to $(\alpha\ket{0}_s\ket{0}_p\ket{\psi_0}_r+\beta\ket{1}_s\ket{\text{vacuum}}_p\ket{\psi_1}_r)$. We then apply a $\pi$-pulse of $\Omega_1$(t) to swap the two ground states, transforming the system into the state $(\alpha\ket{1}_s\ket{0}_p\ket{\psi_0}_r+\beta\ket{0}_s\ket{\text{vacuum}}_p\ket{\psi_1}_r)$. Following another $\pi$-pulse of $\Omega_2$(t) and spontaneous emission, the state of the system is transformed into the state $(\alpha\ket{1}_s\ket{0}_p\ket{\psi_0}_r+\beta\ket{0}_s\ket{1}_p\ket{\psi_1}_r)$. Finally, we apply a $\sigma_x$ measurement on the emitter to disentangle it from the photons as described in the case of Fig.~\ref{cavity}(a). To implement the delayed feedback and the CZ gate, we employ a pair of phase-tunable Mach-Zehnder interferometers to guide the photonic wavepacket in the later time bin ($\ket{1}_p$) back to the cavity. Upon re-scattering, the $\ket{1}_p$ photonic wavepacket picks up a $\pi$ phase shift when the emitter is in the state $\ket{1}_p$, but no phase shift when the emitter is in the state $\ket{0}_p$, which realizes a CZ gate between the photon and the emitter qubit. We reset the phase between the two Mach-Zehnder interferometers to guide this wavepacket to the output when it arrives for the second time, so that the CZ gate is only applied once on each photon. We note that the feedback line introduces an extra delay between the earlier and later time-bin wavepackets of the same photonic qubit. This delay can be counteracted by sending the output to another phase-tunable Mach-Zehnder interferometer setup.

\section{Requirement on cooperativity}

\begin{figure}[htb]
\centering
\includegraphics[width=0.4\columnwidth]{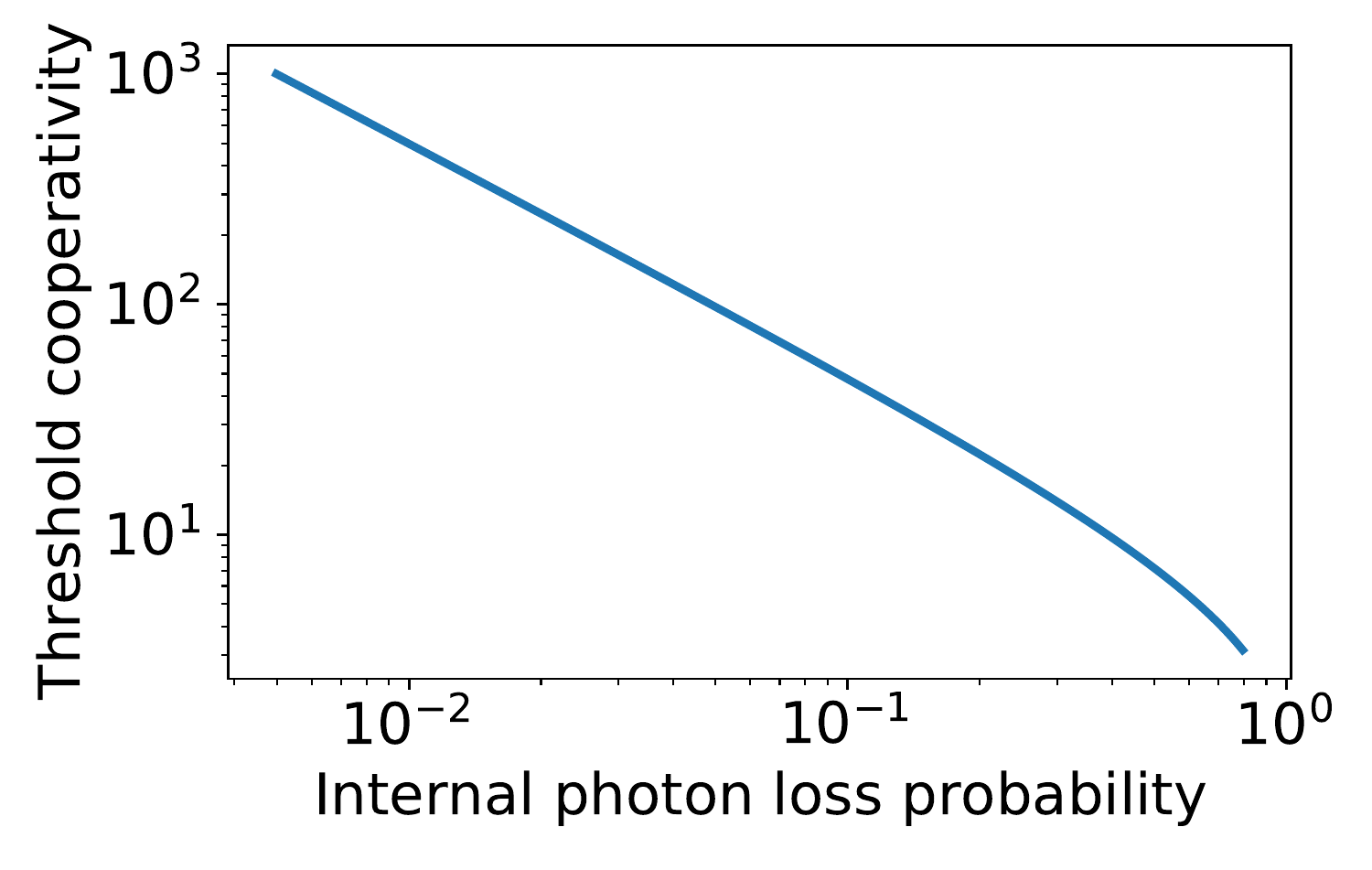}
\caption{The threshold cooperativity as a function of the internal photon loss probability, assuming all other loss channels are absent.}
\label{cooperativity}
\end{figure}

Supplemental Section S1 considers experimental implementations of our scheme using an ideal cavity QED system with an infinite cooperativity. In practice the cooperativity cannot be infinitely large. A finite cooperativity will lead to two loss channels: 1) emission of photons to undesired modes other than the cavity mode during the E gate, and 2) scattering of photons into undesired modes during the CZ gate. Fortunately, both loss channels belong to the internal photon loss that can be corrected using the tree encoding. Given an internal photon loss budget, we can calculate the threshold value of the cooperativity by assuming all other loss channels are absent. Figure~\ref{cooperativity} shows the calculated threshold cooperativity as a function of the internal photon loss probability. For an internal photon loss probability of 0.01 (as we assumed in Fig. 3(b) of the main text), the threshold value of cooperativity is 500.
Such cooperativity values have been realized in quantum dot based cavity QED systems~\cite{Ota:2018aa,Najer:2019aa}. It may also be achievable with color center~\cite{Bhaskar:2020aa} or neutral atom~\cite{RevModPhys.87.1379} based cavity QED systems upon reasonable improvement ($5\sim 10\times$) of the state-of-the-art.

\section{Calculation of effective error probability}

\subsection{Effective error probability with a perfect emitter qubit}

The blue circles in Fig. 3(a) show the effective error probability $\varepsilon_{\text{eff}}$ as a function of the tree size. Recall that $\varepsilon_{\text{eff}}$ is defined as the probability that an error occurs when performing a single qubit measurement on the tree-encoded logic qubit. With a perfect emitter qubit, the effective error probability is equal to the effective loss probability of the logic qubit, $\varepsilon_{\text{loss}}$, given by~\cite{PhysRevLett.97.120501}
\begin{equation}
\begin{aligned}
\varepsilon_{\text{eff}}=\varepsilon_{\text{loss}}=1-[(1-\varepsilon+\varepsilon R_1)^{b_0}-(\varepsilon R_1)^{b_0}](1-\varepsilon+\varepsilon R_2)^{b_1},
\end{aligned}
\label{s1}
\end{equation}
where $\{b_0,\cdots,b_{d-1}\}$ are the branching parameters of the tree, $d$ is the depth of the tree, and $R_l$ is the success probability of performing an indirect $\sigma_z$ measurement on a level-$l$ photon, given by
\begin{equation}
\begin{aligned}
R_l=1-[1-(1-\varepsilon)(1-\varepsilon+\varepsilon R_{l+2})^{b_{l+1}}]^{b_l}\quad\text{for } l<d,
\end{aligned}
\end{equation}
and $R_{d}=R_{d+1}=0$, $b_d=0$.

\subsection{Effective error probability under limited coherence time of the emitter qubit}

The orange squares in Fig. 3(a) show the effective error probability as a function of the tree size when the emitter qubit has a limited coherence time. In this scenario, some photons may experience a phase error during the generation process, which will lead to an uncorrectable error in the tree-encoded logic qubit. We define the effective error probability as $\varepsilon_{\text{eff}}=1-(1-\varepsilon_{\text{loss}})(1-\varepsilon_{\text{coh}})$, where $\varepsilon_{\text{loss}}$ is the effective loss probability of the logic qubit, and $\varepsilon_{\text{coh}}$ is the decoherence probability of the emitter qubit during the tree generation process. The expression of $\varepsilon_{\text{loss}}$ is given by Eq.~(\ref{s1}). The expression of $\varepsilon_{\text{coh}}$ is given by $\varepsilon_{\text{coh}}=1-e^{-t_{\text{min}}/t_{\text{coh}}}$, where $t_{\text{min}}$ is the minimum time during which the emitter needs to stay coherent, and $t_{\text{coh}}$ is the coherence time of the emitter. Note that $t_{\text{min}}$ is smaller than the total time of the generation process. This is because the emitter is disentangled with all the photons following each E gate. Therefore, a dephasing event occurring during the time interval between an E gate and a subsequent E or CZ gate will not affect the coherence of the generated photons. We can calculate $t_{\text{min}}$ as $t_{\text{min}}=\sum_{l=0}^dt_l$, where $t_l$ is the total time the emitter needs to stay coherent during the generation of all the photons in level-$l$ of the tree, given by
\begin{equation}
\begin{aligned}
t_l=\left\{
        \begin{array}{lr}
        [(b_l-1)\Delta t_{l+1}+2t_{\text{ph}}]n_l, &\quad 0\leq l\leq d-1, \\
        t_{\text{ph}}n_l, &\quad l=d,
        \end{array}
\right.
\end{aligned}
\end{equation}
where $n_l$ is the number of photons in level-$l$, and $\Delta t_l$ is the time interval between two neighboring level-$l$ photons, given by
\begin{equation}
\begin{aligned}
\Delta t_l=b_l\Delta t_{l+1}\quad\text{for $1\leq l\leq d-2$},
\end{aligned}
\end{equation}
and $\Delta t_{d-1}=(b_{d-1}+1)t_{\text{ph}}$, $\Delta t_d=t_{\text{ph}}$.

\subsection{Calculation of the optimal effective error probability}

In Fig. 3(b) of the main text, we calculate the optimal effective error probability $\varepsilon_{\text{eff}}$ as we vary both the single-photon loss probability $\varepsilon$ and the coherence time of the emitter qubit $t_{\text{coh}}$. We normalize the coherence time in terms of the inverse of the bandwidth of the CZ gate, $\gamma_R$. In this calculation, we assume $\gamma_B/\gamma_R=0.0014$ and $\gamma_Bt_{\text{ph}}=6.2$, where $\gamma_B$ is the bandwidth of the photon emitted by the E gate, and $t_{\text{ph}}$ is the time allocated to a single photonic qubit. The first assumption $\gamma_B/\gamma_R=0.0014$ ensures that each incident photon is quasi-monochromatic to the CZ gate, such that the infidelity of the CZ gate satisfies $\varepsilon_{\text{CZ}}<10^{-5}$, two orders of magnitude smaller than the targeted value of $\varepsilon_{\text{eff}}$ ($10^{-3}$). The second assumption $\gamma_Bt_{\text{ph}}=6.2$ ensures that the overlap of the photonic wavepackets between the earlier and later time bins satiesfies $\varepsilon_{\text{ol}}<10^{-5}$, again two orders of magnitude smaller than the targeted value of $\varepsilon_{\text{eff}}$. We now verify these assumptions with explicit calculations.

We first calculate the infidelity of the CZ gate when the incident photon has a finite bandwidth $\gamma_B$. We calculate the gate infidelity when both the emitter qubit and the incident photon is in the superposition state, given by
\begin{equation}
\begin{aligned}
\ket{\psi_i}=\frac{1}{2}\int_{-\infty}^{\infty}f(t)(\ket{0}_s\ket{0}_p+\ket{0}_s\ket{1}_p+\ket{1}_s\ket{0}_p+\ket{1}_s\ket{1}_p)dt,
\end{aligned}
\end{equation}
where $f(t)$ is the temporal profile of the incident photon, satisfying $\int_{-\infty}^{\infty}|f(t)|^2dt=1$. The gate infidelity is given by
\begin{equation}
\begin{aligned}
\varepsilon_{\text{CZ}}=1-|\braket{\psi_{fi}|\psi_f}|^2,
\end{aligned}
\label{cz}
\end{equation}
where $\ket{\psi_{fi}}$ and $\ket{\psi_f}$ are the states of the system following an ideal and actual CZ gate respectively, given by
\begin{equation}
\begin{aligned}
\ket{\psi_{fi}}=\frac{1}{2}\int_{-\infty}^{\infty}f(t)(\ket{0}_s\ket{0}_p+\ket{0}_s\ket{1}_p+\ket{1}_s\ket{0}_p-\ket{1}_s\ket{1}_p)dt,
\end{aligned}
\label{psi_fi}
\end{equation}

\begin{equation}
\begin{aligned}
\ket{\psi_f}=\frac{1}{2}\int_{-\infty}^{\infty}[g_{00}(t)\ket{0}_s\ket{0}_p+g_{01}(t)\ket{0}_s\ket{1}_p+g_{10}(t)\ket{1}_s\ket{0}_p+g_{11}(t)\ket{1}_s\ket{1}_p]dt.
\end{aligned}
\label{psi_f}
\end{equation}
Here, $g_{ij}(t)$ is the temporal profile of the transmitted photonic wavepacket when the initial states of the emitter and the photon are $\ket{i}_s$ and $\ket{j}_p$ respectively $(i, j \in\{0,1\})$. Substituting the expressions of $\ket{\psi_{fi}}$ and $\ket{\psi_{f}}$ into Eq.~(\ref{cz}), we obtain that
\begin{equation}
\begin{aligned}
\varepsilon_{\text{CZ}}=1-\left|\frac{1}{4}\int_{-\infty}^{\infty}f^\ast(t)[g_{00}(t)+g_{01}(t)+g_{10}(t)-g_{11}(t)]dt\right|^2.
\end{aligned}
\label{czz}
\end{equation}

To calculate $\varepsilon_{\text{CZ}}$, we assume that the incident photonic wavepacket has a Gaussian temporal profile, given by
\begin{equation}
\begin{aligned}
f(t)=\frac{1}{\sqrt{2\pi}}\sqrt{\frac{1}{\gamma_B\sqrt{\pi}}}\int_{-\infty}^{\infty}e^{-\frac{\omega^2}{2\gamma_B^2}}e^{i\omega t}d\omega.
\end{aligned}
\label{gaussian}
\end{equation}
As suggested in Ref.~\cite{Pichler:2017aa}, the Gaussian pulse shape is more robust against distortion following the emitter-photon CZ gate when compared with an exponential pulse. Experimentally, we can generate the Gaussian pulsed emission by employing a Raman emission process with the help of the ancillary level $\ket{g_2}$. Such Raman emission processes have been experimentally demonstrated using trapped Rb atoms~\cite{PhysRevLett.85.4872,PhysRevLett.89.067901}, semiconductor quantum dots~\cite{Sweeney:2014aa}, and silicon-vacancy color centers in diamond~\cite{PhysRevLett.121.083601}. 

We also assume that the CZ gate operates perfectly in the basis $\{\ket{0}_s\ket{0}_p,\ket{0}_s\ket{1}_p,\ket{1}_s\ket{0}_p\}$, such that $g_{00}(t)=g_{01}(t)=g_{10}(t)=f(t)$. This assumption is valid since the emitter and the photon are decoupled in these bases. To calculate $g_{11}(t)$, we use the fact that the atom-waveguide system is linear with single-photon input. Thus $g_{11}(t)$ is given by
\begin{equation}
\begin{aligned}
g_{11}(t)=\sqrt{\frac{1}{2\pi \gamma_B\sqrt{\pi}}}\int_{-\infty}^{\infty}t(\omega)e^{-\frac{\omega^2}{2\gamma_B^2}}e^{i\omega t}d\omega,
\end{aligned}
\label{g11}
\end{equation}
where $t(\omega)$ can be calculated based on input-output relation~\cite{walls2007quantum}, given by
\begin{equation}
\begin{aligned}
t(\omega)=\frac{a_{\text{out}}}{a_{\text{in}}}=\frac{\omega-i\frac{\gamma_R}{2}}{\omega+i\frac{\gamma_R}{2}}.
\end{aligned}
\label{t}
\end{equation}

Substituting Eq.~(\ref{gaussian})-(\ref{t}) into Eq.~(\ref{czz}), we obtain that
\begin{equation}
\begin{aligned}
\varepsilon_{\text{CZ}}=2(\gamma_B/\gamma_R)^2+\mathcal{O}\left[(\gamma_B/\gamma_R)^4\right].
\end{aligned}
\end{equation}
When $\gamma_B/\gamma_R=0.0014$, we obtain that $\varepsilon_{\text{CZ}}=4\times 10^{-6}$.

We next calculate the overlap between photonic wavepackets in the earlier and later time bins, given by
\begin{equation}
\begin{aligned}
\varepsilon_{\text{ol}}=\int_{t_{\text{ph}}/2}^\infty |f(t)|^2dt.
\end{aligned}
\end{equation}
Assuming a Gaussian temporal profile of photonic wavepacket given by Eq.~(\ref{gaussian}), we obtain that 
\begin{equation}
\begin{aligned}
\varepsilon_{\text{ol}}=\frac{1-\text{erf}\left(\frac{\gamma_B t_{\text{ph}}}{2}\right)}{2},
\end{aligned}
\end{equation}
where $\text{erf}(x)$ is the Gaussian error function. For $\gamma_B t_{\text{ph}}=6.2$, we obtain that $\varepsilon_{\text{ol}}=6\times 10^{-6}$.

\section{Error and fidelity analysis under experimental imperfections}

\subsection{Single-photon logic error}

The analysis in the main text mainly focused on single-photon loss errors, where no detection event is registered for a subset of photons in the tree state. However, the generated state may also suffer from single-photon logic errors. In this case a subset of photons in the tree successfully make it to the detector, but the measured result by the detector is opposite to the expected value for an error-free tree state. These errors can arise from finite indistinguishability of the emitter, the overlap between the wavepackets in the two time-bins, the detection error caused by the detector dark counts, etc.

\begin{figure}[htb]
\centering
\includegraphics[width=0.7\columnwidth]{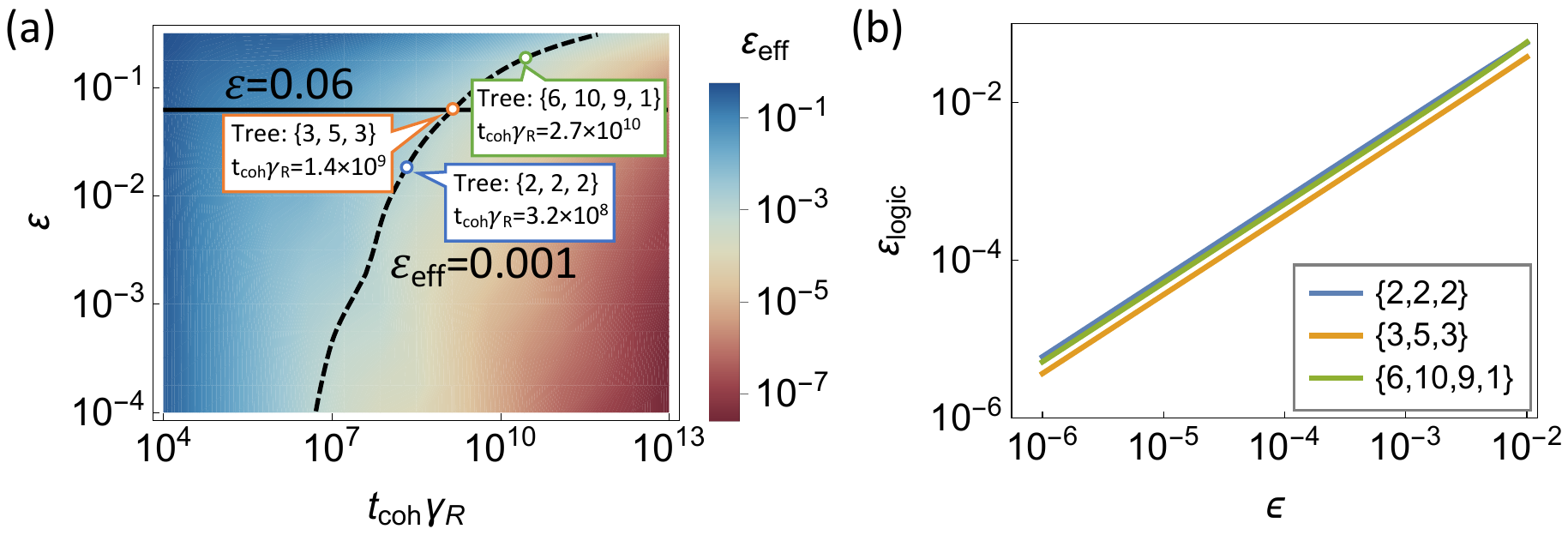}
\caption{(a) Reproduction of Fig. 3(b) of the main text. In this figure, we label the three optimized tree structures we choose for logic error analysis, whose branching parameters are $\{2,2,2\}$, $\{3,5,3\}$, and $\{6,10,9,1\}$, respectively. (b) The logic error probability $\varepsilon_{\text{logic}}$ of the tree-encoded logic qubit as a function of the single-photon logic error $\epsilon$. Here, we assume the single-photon logic error remains the same for all measurement bases. The logic error probability is calculated in the basis that gives the largest error. We assume the loss is absent in this analysis.}
\label{logical}
\end{figure}

An important question is whether the logic error in each single photon can lead to a large accumulated logic error in the tree-encoded logic qubit. Fortunately, this is not the case as shown by previous analysis~\cite{Azuma:2015aa,hilaire2020resource}, since the tree-encoding allows correction of the single-photon logic errors through a majority voting process. Here we apply the same analysis to three specific tree structures, with branching parameters $\{2,2,2\}$, $\{3,5,3\}$, and $\{6,10,9,1\}$ respectively. These three branching parameters correspond to three optimized tree structures that enable $\varepsilon_{\text{eff}}=0.001$ in Fig. 3(b) (dashed line) of the main text. Figure~\ref{logical}(a) is a reproduction of Fig. 3(b) of the main text. We have labeled the operation condition of the three tree structures in Fig.~\ref{logical}(a).

Figure~\ref{logical}(b) shows the logic error probability $\varepsilon_{\text{logic}}$ of the tree-encoded logic qubit as a function of the single-photon logic error $\epsilon$. Note that the single-photon logic error can be different depending on the measurement basis. For example, the finite indistinguishability of the emitter will not create a logic error in the $\sigma_z$ basis measurement, but it will cause an error in the $\sigma_x$ basis measurement when we need to interfere the wavepackets that belong to the two different time-bins. On the other hand, the detector dark count will cause an uniform logic error probability for all measurement bases. Here we assume the single-photon logic error remains the same for all measurement bases. This assumption leads to the worst possible performance. We also provide the error of the tree-encoded logic qubit in the basis that gives the largest error probability. From Fig.~\ref{logical}(b), we can see that the logic error of the logic qubit scales linearly with the single-photon logic error, with a value that never exceeds 6 times of the single-photon logic error for the tree structures considered here. If we want to maintain the logic error of the logic qubit below $10^{-4}$ (10 times smaller than the targeted effective error rate in the analysis of the main text), then we need a single-photon logic error probability of $\sim 10^{-5}$. Since the errors caused by both the time-bin overlap (calculated to be $6\times 10^{-6}$ in Sec. S3) and the detector dark count ($<10^{-9}$ with superconducting nanowire single-photon detectors) are lower than $10^{-5}$, this error requirement suggests that we need the emitter distinguishability to be $<10^{-5}$ between the emission into the two adjacent time-bins.

\subsection{Fidelity analysis under other experimental imperfections}

\begin{figure}[htb]
\centering
\includegraphics[width=0.77\columnwidth]{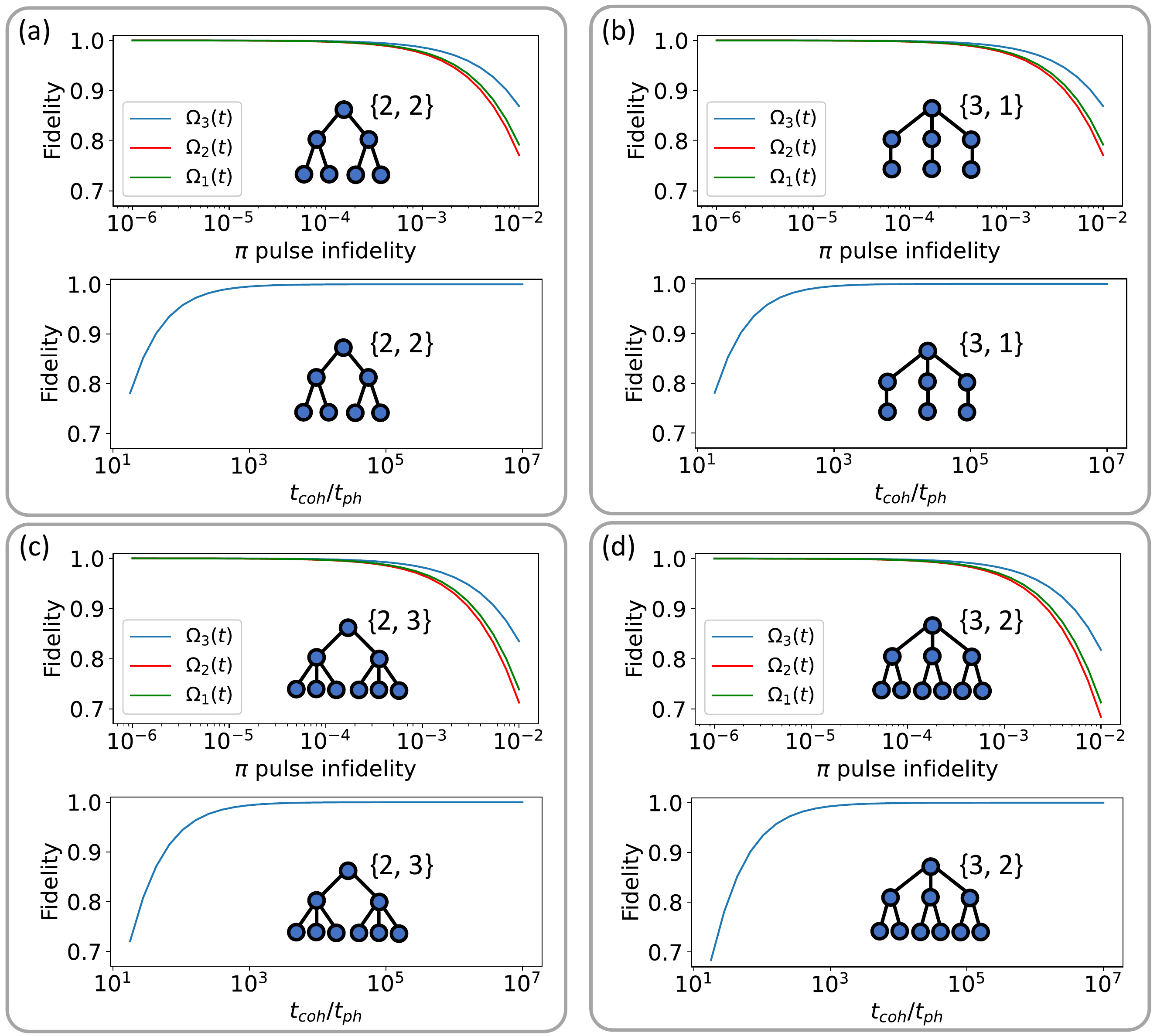}
\caption{The fidelity of the generated state as a function of $\pi$ pulse infidelity (top figure of each panel) and the coherence time of the emitter qubit (bottom figure of each panel) for trees with branching parameters $\{2,2\}$(a), $\{3,1\}$(b), $\{2,3\}$(c), and $\{3,2\}$(d). In the top panels, we use three colors, green, red, and blue, to represent the imperfect $\pi$ rotation pulses of $\Omega_1(t)$, $\Omega_2(t)$, and $\Omega_3(t)$ respectively. The effect of an imperfect $\pi$ pulse is modeled by a quantum depolarizing channel of the form $\Delta_\lambda(\rho)=(1-\frac{3\lambda}{2})\sigma_x\rho \sigma_x+\frac{\lambda}{2}(\rho+\sigma_y\rho\sigma_y+\sigma_z\rho\sigma_z)$, where $\lambda$ is the $\pi$ pulse infidelity and $\sigma_{x,y,z}$ are Pauli matrices.}
\label{fidelity}
\end{figure}

Experimental realization of our scheme may suffer from other imperfections that cannot be categorized into the single-photon loss or single-photon logic errors. These imperfections include the imperfect $\pi$ pulses in the implementation of the E gate and the CZ gate, and finite coherence time of the emitter. It remains an open question whether these imperfections are quantum error correctable with the tree-encoding. To quantitatively understand the effect of these experimental imperfections, we calculate the fidelity of the generated state as we introduce each experimental imperfection. We define the fidelity as $F=\bra{\psi_{\text{ideal}}}\rho\ket{\psi_{\text{ideal}}}$, where $\ket{\psi_{\text{ideal}}}$ is the ideal tree state, and $\rho$ is the density matrix of the generated photons under experimental imperfections. Figure~\ref{fidelity} shows the calculated fidelity as a function of different parameters for a few tree structures. Note that we can only perform fidelity analysis for tree states that contain a small number of photons ($<11$) since the dimension of the density matrix grows exponentially with the number of photons in the tree. This fidelity analysis will be particularly relevant for initial proof-of-concept experimental demonstration of our tree generation scheme. For example, if we want to generate a tree with branching parameters $\{2,2\}$ while achieving a fidelity exceeding 0.77, we need the $\pi$ pulse fidelity exceeding 0.99. This condition is achievable with silicon-vacancy color centers~\cite{PhysRevLett.119.223602}, and can also be achieved upon reasonable improvement of the qubit rotation fidelity of trapped atoms~\cite{Reiserer:2014aa}, and semiconductor quantum dots~\cite{De-Greve:2012aa}.

\section{Generalization of the proposed scheme}

\begin{figure}[htb]
\centering
\includegraphics[width=0.65\columnwidth]{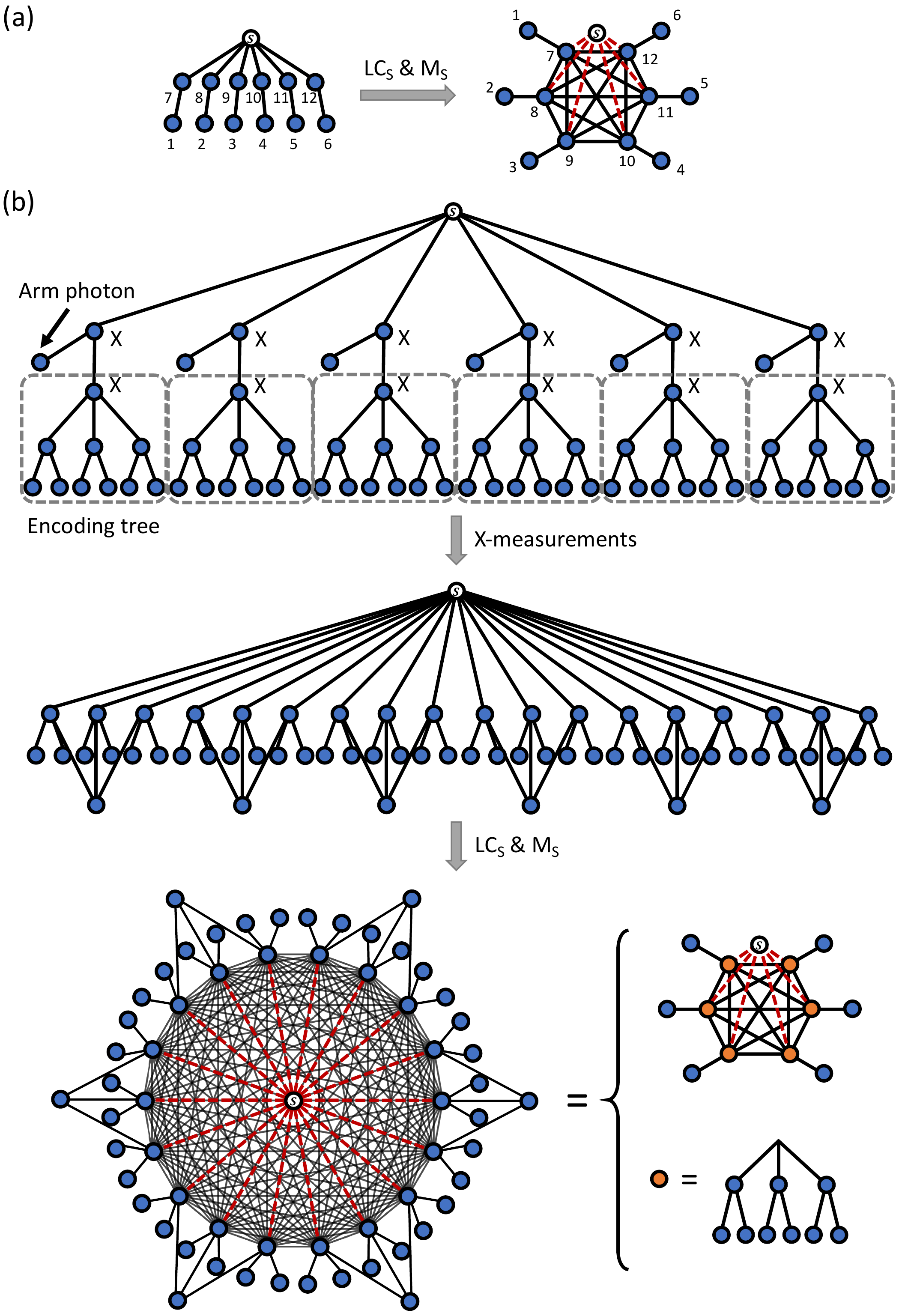}
\caption{Schematic sequence to generate a repeater graph state with 6 core photons ($N=6$) (a) and a tree-encoded repeater graph state with each core logic qubit encoded by a $\{3,2\}$ tree (b). The red dashed lines mean that the corresponding entanglements vanish after the $\sigma_z$ measurement on the emitter.}
\label{rgs}
\end{figure}

Our protocol can be broadly applicable to the generation of a variety of photonic cluster states with tree-type entanglement structures. One example is the repeater graph state, which is a useful multi-photon entangled state in the implementation of an all-optical quantum repeater~\cite{Azuma:2015aa}. Figure~\ref{rgs}(a) illustrates the schematic sequence to generate a repeater graph state with 6 core photons ($N=6$). The repeater graph state can be constructed from a 3-layer tree with branching parameters $b_0=N$ and $b_1=1$, with the root qubit being the emitter. We then apply a local complementation operation on the emitter, which only includes single-qubit gates on the emitter $\left(e^{i\frac{\pi}{2}\frac{Y+Z}{\sqrt{2}}}\right)$ and all the $N$ core photons $\left(e^{i\frac{\pi}{2}\frac{X+Y}{\sqrt{2}}}\right)$~\cite{PhysRevA.69.062311,PhysRevX.7.041023}. After this step, the $N$ core photons will be completely connected. We finally apply a $\sigma_z$ measurement on the emitter to detach it from the graph. The mathematical description of the generation process is as follows:
\begin{equation}
\begin{aligned}
\ket{\psi}_{\text{RGS}}=M_S\left(e^{i\frac{\pi}{2}\frac{Y+Z}{\sqrt{2}}}\right)_{S}\left[\prod_{k=1}^N \left(e^{i\frac{\pi}{4}Y}\right)_S\left(e^{i\frac{\pi}{2}\frac{X+Y}{\sqrt{2}}}\right)_{N+k}E_{S,N+k} CZ_{S,k}\right]\left[\prod_{j=1}^N\left(e^{i\frac{\pi}{4}Y}\right)_SE_{S,j}\right]\ket{+}_s,
\end{aligned}
\end{equation}
where $M_S$ represents the $\sigma_z$ measurement on the emitter.

Our protocol can also be generalized to generate the tree-encoded repeater graph states. These states combine the tree-entanglement with the repeater graph state, which are more resilient to both the loss and the decoherence errors~\cite{PhysRevA.95.012304}. The bottom panel of Fig.~\ref{rgs}(b) shows an example of a tree-encoded repeater graph state. The state consists of 6 core qubits and 6 arms ($N=6$), where each core qubit (represented by the orange circles) is logically encoded by a tree with branching parameters $\{3,2\}$. To generate this state requires a similar process with the repeater graph state. Figure~\ref{rgs}(b) illustrates the state generation process. We start with a more complex tree state [as shown in the top panel of Fig.~\ref{rgs}(b)]. We then perform $\sigma_x$ measurements on all the level-1 photons and the root photons of the encoding trees. The two $\sigma_x$ measurements on each adjacent photon pair remove the two photons and form direct bonds among their neighbors~\cite{PhysRevLett.97.120501}. Finally, a local complementation operation followed by a $\sigma_z$ measurement on the emitter qubit completes the generation of the tree-encoded repeater graph state.

In all cases, we rely on only single-photon emission and scattering from a single quantum emitter to generate the whole entangled cluster state, which is both fast and resource-efficient compared with all previous protocols. This capability demonstrates the versatility of our protocol in the generation of a large range of photonic cluster states that possess tree-type entanglement.

\providecommand{\noopsort}[1]{}\providecommand{\singleletter}[1]{#1}%